\documentclass[usenatbib]{mn2e}
\usepackage{graphicx}
\newcommand{\text}[1]{\quad\mbox{#1}\quad}

\newcommand{\aj}{AJ}
\newcommand{\apj}{ApJ}
\newcommand{\apjs}{ApJS}
\newcommand{\apjl}{ApJ}
\newcommand{\mnras}{MNRAS}

\newcommand{\aap}{A\&A}

\newcommand{\sub}[1]{_{\mbox{\tiny #1}}}

\title{Hard X-ray Bursts from Collapse of the Super Massive Stars} 
\author[M.V.~Barkov ]{Maxim V.~Barkov,$^{1,2,3}$
\thanks{E-Mail:
bmv@mpi-hd.mpg.de (MVB)}\\
$^{1}$Space Research Institute, 84/32 Profsoyuznaya Street, Moscow
117997, Russia\\
$^{2}$Department of Applied Mathematics, The University of Leeds,
Leeds, LS2 9JT, UK\\
$^{3}$Max-Planck-Institut f\"ur Kernphysik, Saupfercheckweg 1, 69117 Heidelberg, Germany}

\begin{document}
\date{Received/Accepted}
\maketitle

\begin{abstract}
The very first stars in the Universe can be very massive, up to $10^6M_\odot$. 
They would leave behind massive black holes that could act as seeds 
for growing super massive black holes of active galactic nuclei. Given the anticipated fast
rotation such stars would end their live as super massive collapsars and drive powerful 
magnetically-dominated jets. In this paper we investigate the possibility of observing 
the bursts of high-energy emission similar to the Long Gamma Ray Bursts associated with 
normal collapsars.  We show that during the collapse of supercollapsars, the 
Blandford-Znajek mechanism can produce jets as powerful as 
few$\times10^{51}$erg/s and release up to $10^{56}$erg of the black hole rotational 
energy. Due to the higher intrinsic time scale and higher redshift the initial bright 
phase of the burst can last for about $10^5$ seconds whereas the central engine would 
remain active for about 10 days.  
Due to the high redshift the burst spectrum is expected to be soft, with the    
spectral energy distribution peaking at around 60keV. The peak total flux density is 
relatively low, few$\times 10^{-7}\mbox{erg}\, \mbox{cm}^{-2} \mbox{s}^{-1}$, but not prohibitive.   
The such events should be rear 0.03 year$^{-1}$, the observations needs long term program and could be done in future.     

\end{abstract}

\maketitle

\section{Introduction}
\label{intro}

The very first stars in universe were borned in the lack of heavy elements.
It could leads to born of the super  massive stars (SMS)  \citep{bcl02,bl03,ss06,dv09}.
Such stars can be as massive as $1000 M_{\odot} < M < 50000 M_{\odot}$.

The stars with masses about $ 1000 M_{\odot}$, which will be referred to as Very Massive Stars (VMSs) are expected to collapse into black holes with very little mass loss \citep{F01}.   
They would leave behind massive black holes (MBHs), which could play the role of seeds 
for the super massive black holes (SMBHs) of Active Galactic Nuclei (AGNs). 
The collapse of VMS were discussed recently in the paper \citep{KB09c}.

Even more massive $3\times 10^4 M_{\odot} < M < 10^6 M_{\odot}$
stars, which will be referred to as Super Massive Stars (SMSs) could be formed in more massive dark 
matter haloes with total mass $M\simeq 10^8M_{\odot}$ collapsed at $z\simeq10$ 
\citep{bl03,BVR06}. This stars do not reach the instability limit mass which is near $10^6 M_{\odot}$ and can be formed \citep{hf63,zn65,bzn67,w69}. The collapse of SMS can provide an alternative way of producing SMBHs.

 

There are two crucial differences between a normal collapsar and a supercollapsar. 
One is that instead of a proto-neutron star of solar mass the supercollapsars
develop proto-black holes of tens of solar masses,  
within which the neutrinos from electron capture are trapped \citep{F01,Su07}. 
The other is that the accretion disks of supercollapsars are far too large and cool 
for the neutrino annihilation mechanism. This has already been seen in the numerical 
simulations of supercollapsar with mass $M=300M_\odot$ \citep{F01}. Utilising  
the study of hyper-accreting disks by \citet{bel08} we find that at best the  
rate of heating due to this mechanism is 
\begin{equation}
\dot{E} \simeq 6\times10^{43} \dot{M}_0^{9/4}  M_{h,6}^{-3/2}  
 \mbox{erg s}^{-1},
\end{equation}    
where $\dot{M}$ is the accretion rate and $M_h$ is the black hole mass. 
(Here and in other numerical estimates below we use the following 
notation:  $\dot{M}_k$ is the mass accretion rate measured in the units of 
$10^k M_{\odot} \mbox{s}^{-1}$ and $M_k$ is the mass measured in the units of 
$10^kM_\odot$.) Such low values  
have lead \citet{F01} to conclude 
that the magnetic mechanism is the only candidate for producing GRB jets 
from supercollapsars. In the following we analyse one particular version 
of the mechanism where the jets are powered by the rotational energy of the 
black hole via the Blandford-Znajek process \citep{BZ77,BK08}.

This paper is an extension our previous work \citep{KB09c} here we investigate 
properties of collapse of VMS. In this paper we estimate temporal structure of the SMS 
collapse and predict the observational evidence of this process.

\section{Physical model}
\label{pmodel}

\begin{figure*}
\includegraphics[width=74mm,angle=0]{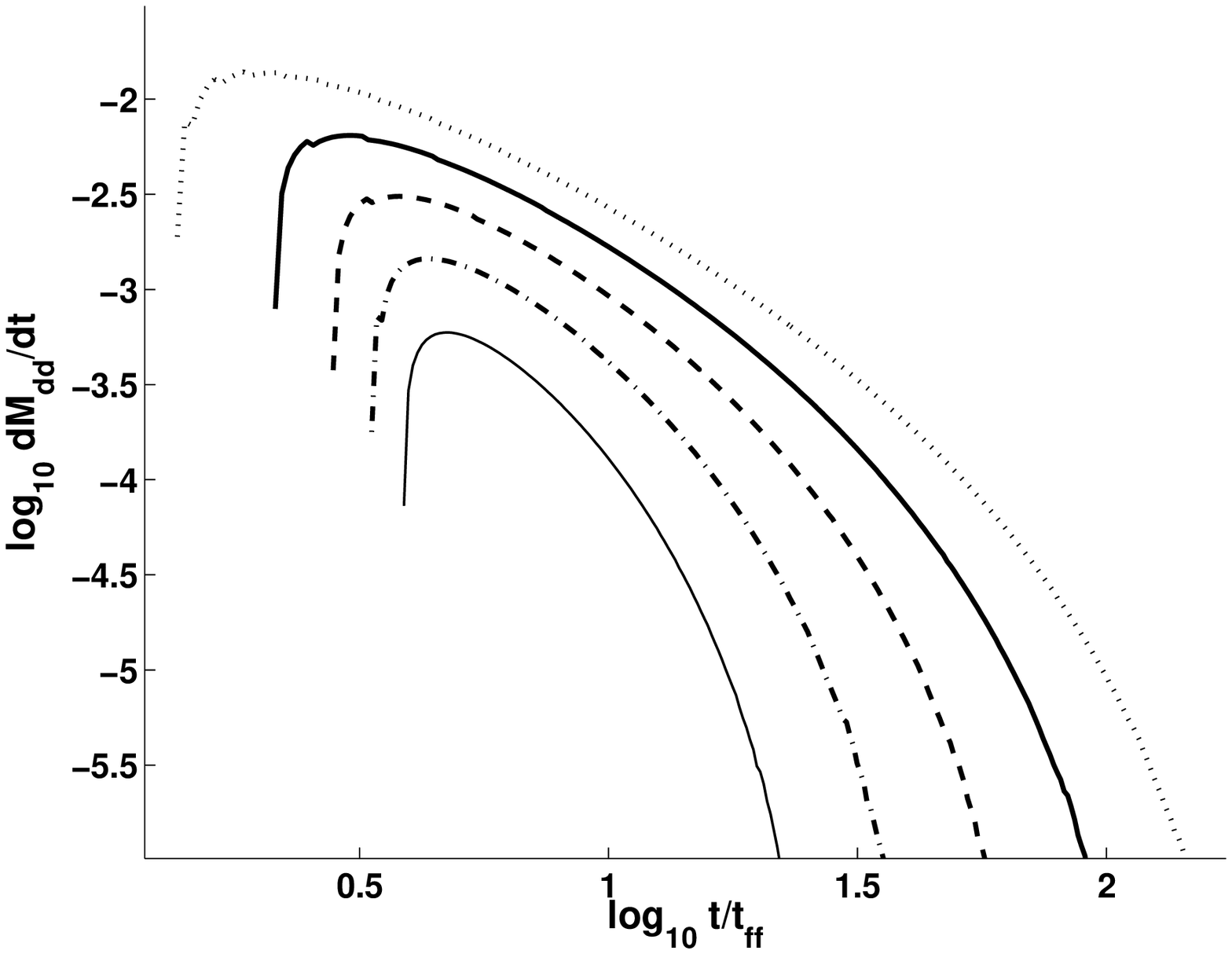}
\includegraphics[width=74mm,angle=0]{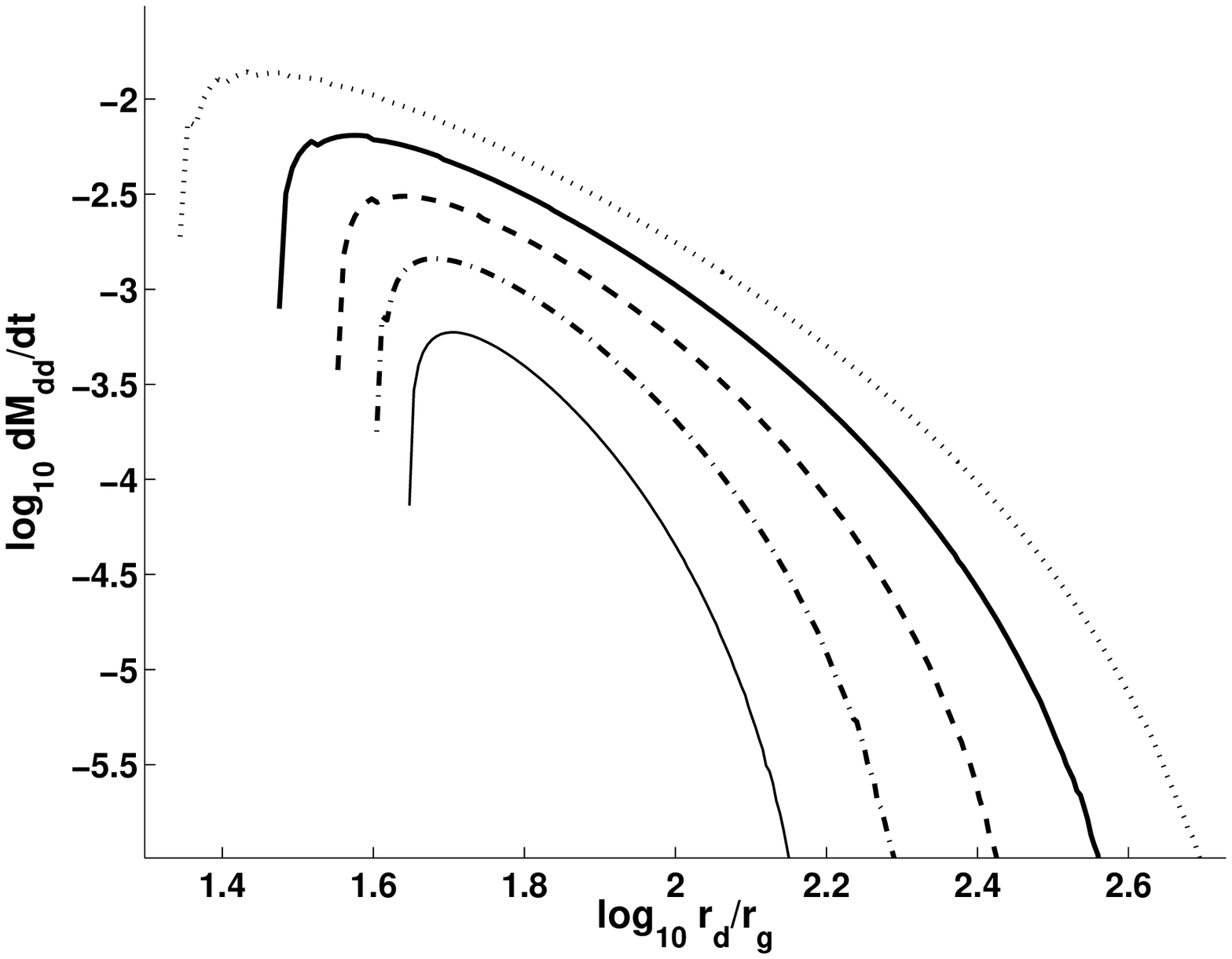}
\caption{Accretion disck rate depence on time (left panel) and effective accretion 
disk radius (right panel). The plots in the units $M_{s}/t_{ff} = \frac{2.3\times 10^{-38}c^3}{G}=9.4 M_{\odot} s^{-1}$, $\alpha=1/30$.
doted line is for J = 0.97, thick solid -- 0.82, dashed --  0.69, dot-dashed -- 0.58, thin solid -- 0.49} 
\label{mdot}
\end{figure*}

Lets look for this problem. SMS should be fully convective and have 
rigid rotation. In our simple calculations we neglect aspherical
shape of fast rotation SMS. 

We assume the radius SMS before collapse as 640 $GM/c^2$ \citep{zn71,st83,shib02}.
We can derive the undimensional critical parameter of rotation for SMS with $3\times10^4 M_{\odot}<M< 10^6 M_{\odot}$ 
as $J_s<(M/10^6 M_{\odot})^{1/4}$.
Numerical simulations of SMS collapse show formation of 
massive BH and accretion disk .
Follow the works \citep{shib02,bk09} we know spin parameter of 
BH and mass of accretion disk. 

In sake of simplicity we use model of the advection dominated 
accretion flow (ADAF) \citep{ny94}.


Assuming gas pressure to dominate, we write the pressure $P = \rho c_s^2$ where
$\rho(R)$ is the height-averaged density and $c_s(R)$ is the isothermal 
sound speed,  we also put $\gamma = 4/3$.
\begin{equation}
v_k=\sqrt{\frac{GM}{R}}, \qquad v_r \approx -\frac{3 \alpha}{7}v_k, 
 \qquad c_s^2 \approx \frac{2}{7}v^2_k,
\label{vr}
\end{equation}
here $v_r$ is radial velocity, $\alpha$ is a constant \citep{shs73}. 
Vertical scale height is $H \sim R c_s/v_k$. It is easy to see
that disk accretion time can be expressed as
\begin{equation}
t_d = -\frac{2R_d}{3v_r} \approx \frac{14}{9\alpha}\sqrt{\frac{R_d^3}{GM_s}}
\label{dat}
\end{equation}
Lets assume accretion disk mass as $M_d = D  M_s$, the accretion rate is 
determined by defuse time from maximum disk mass radius 
 \begin{equation}
\dot{M} \approx \frac{M_d}{t_d} \approx \frac{9D\alpha}{14}\left(\frac{GM_s^3}{R_d^3}\right)^{1/2}.
\label{md}
\end{equation}
Using continuity equation $\dot{M} = -4\pi RHv_r\rho$ and exp. (\ref{md}), get
 \begin{equation}
\rho \approx \frac{3\sqrt{14}}{16\pi}\frac{DM_s}{R^{3/2}R_d^{3/2}}.
\label{rho}
\end{equation} 
combine (\ref{vr}) and (\ref{rho}) we get
 \begin{equation}
P \approx \frac{3\sqrt{14}}{56\pi}\frac{DGM_s^2}{R^{5/2}R_d^{3/2}}.
\label{p}
\end{equation}
magnetic field pressure should be the same order as gas one $B_c^2 / 8\pi \approx P$,
follow the work \cite{tp96} we can estimate large scale poloidal magnetic
field $B_p \approx B_c H/R$, so  
 \begin{equation}
B_p \approx \frac{{3^{1/2}}2^{3/4}}{7^{3/4}}\frac{\sqrt{DG}M_s}{R^{5/4}R_d^{3/4}}.
\label{b}
\end{equation}



After last marginally stable orbit the large scale magnetic field 
could be accreted to BH with gas from accretion disk, the similar approach 
are described in \cite{rgb06}.

$R_{ms}$ is the radius of the marginally stable orbit \citep{bard72},
\begin{equation}
\begin{array}{l}
R_{ms} = r_g \lbrace 3+Z_2\mp [(3-Z_1)(3+Z_1+2Z_2)]^{1/2} \rbrace , \\ 
 Z_1 \equiv 1+(1-a^2/M^2)^{1/3}[(1+a/M)^{1/3}+(1-a/M)^{1/3}],\\ 
Z_2 \equiv (3a^2/M^2+Z_1^2)^{1/2} .
\end{array}
\label{rms}
\end{equation}

To estimate magnetic flux on the horizon of BH we should substitute (\ref{rms}) to
(\ref{b}) so
 \begin{equation}
\Psi = 2\pi B_p R_{ms}^2 \approx 2\pi\frac{{3^{1/2}}2^{3/4}}{7^{3/4}}
\frac{R_{ms}^{3/4}\sqrt{DG}M_s}{R_d^{3/4}}.
\label{psi1}
\end{equation}
the energy realise of Blandford-Znajek mechanism (\citep{BZ77}) could be expressed 
as 
 \begin{equation}
\dot{E}_{BZ} = \frac{1}{6c}\left(\frac{\Psi\Omega}{4\pi}\right)^2 
\label{bz1}
\end{equation}
here $\Omega = \Omega_h/2$ the angular velocity of field lines,
where $\Omega_h=a/2(1+\sqrt{1-a^2})c^3/GM_{BH}$ is angular velocity of BH \citep{k08},
$a=J_{BH}/M_{BH}^2$ is dimensionless speen parameter of BH.
Let us express the current radius and disk radius in $r_g$ units as
$r_{ms}=R_{ms}/r_g$ and $r_d=R_d/r_g$, substituting (\ref{psi1}) to (\ref{bz1}) we get
\begin{equation}
\dot{E}_{BZ} \approx \frac{\sqrt{14}}{3136}\frac{a^2r_{ms}^{3/2}}{(1+\sqrt{1-a^2})^2} 
\frac{c^5}{G} \frac{D}{ r_d^{3/2} } 
\label{bzu1}
\end{equation}
The energy release depend on spin parameter of BH $a$, mass fraction of the disk and radius of the disk.
It is easy to see that all dependence from $0.5<a<1$ is very weak ($0.7\div1.4$) so for our continue estimations we can assume it as 1. The expression (\ref{bzu1}) will be shown as
\begin{equation}
\dot{E}_{BZ} \approx 0.0012 \frac{c^5}{G} \frac{D}{ r_d^{3/2} } 
\label{bzn}
\end{equation}

\section{Results}
\label{Results}


We use the method which is 
described in the paper \cite{bk09} to calculate accretion 
disc fraction and accretion time scales.
The accretion disk which formed for spin parameters of the star
 $0.5<J_s< 0.97$ have longer life time (see fig. \ref{mdot}, left panel) then free-fall time
\begin{equation}
 t_{ff} = \sqrt{\frac{2R^3_s}{9GM_s}}
\label{tff}
\end{equation}
Parameter $D$ depends on star angular momentum.
We interpolate our numerical model for polytropes ($\gamma=4/3$) star 
with good accuracy in the range $0.5<J_s< 0.97$  by simple formula 
\begin{equation}
 log_{10}D = -3J_s^2+7J_s-4.9 
\label{DJ}
\end{equation}.   
For $J_s<0.5$ the accretion disk mass is negligible and $D \ll 1$.

The accretion disk has maximum of matter distribution near 
radius $R_d \approx 25\div50 r_g$  and it is a wick function of $J_s$
(see fig.\ref{mdot} right panel). To simplify our estimations we put $R_d \approx 40 r_g$ 
But accretion could be $3\div10$ times longer (see fig.\ref{mdot} left panel) or you can find 
scaling for self similar solution in the work of \cite{bel08}.

The disk accretion time we can estimate as
\begin{equation}
t_d \sim 1 {M_{s,6}}{\alpha_{-1}}^{-1} \mbox{day}
\label{datn}
\end{equation}

Using the mass accretion rate of fig.(\ref{mdot}) we can check if the neutrino cooling 
needs to be included in the model. Under the conditions of the supercollapsar's 
disk its cooling is dominated by pairs.  
Using the well known equation for this cooling rate \citep[e.g.][]{ykgh01} we can 
compare the cooling time with the accretion time at a given radius. The result is 

\begin{equation}
\frac{t_d}{t\sub{cool}} \simeq 10^{-5} 
  \alpha_{-1}^{-9/4} (R/R_g)^{-13/8}
   \dot{M}_{-1}^{5/4}  M_{s,6}^{\!-3/2}.
\label{cooling}
\end{equation}  
Thus, except for the very 
inner part of the disk, the neutrino cooling is indeed inefficient. 

The complicated magnetic field topology can leads to strong degradation of
energy release, the energy flux could be 10 times lower \citep{bb09}.
Over factor which can decrease magnetic flux on horizon is runaway of magnetic field
if it became to strong \citep{i08}. We will introduce the magnetization parameter in our formula $\beta$ which could be order 0.1. 

Here, we assumed that the whole of the disk is accreted by BH, following the 
original ADAF model. However, it has been argued that this model has to 
be modified via including the disk wind  \citep[ADIOS (Advection Dominated Inflow Outflow 
Solution) model,][]{BB99}, which 
implies a mass loss from the disk and a smaller accretion rate compared to 
Fig.\ref{mdot}. While the arguments for disk wind are very convincing, 
the actual value of mass loss is not well constrained. We can estimate 
the effectiveness of accretion as follow  $\eta \equiv \dot{M} / \dot{M}_{init} \approx (r_{in}/r_{out})^p$, here
$r_{in} \approx r_{ms} \approx 5 r_g $, $r_{out} \approx 100 r_g$ the radius of effective 
nuclear dissociation and $p \approx 0.75$. In the end we have  $\eta \approx 0.1 $.

Using (\ref{bzn}) now we can estimate The maximum posssible energy realise rate is
\begin{equation}
\dot{E}_{BZ} \sim 1.7\times 10^{51} \eta_{-1} \beta_{-1} {D_{-1}} erg  \; s^{-1}, 
\label{bzf}
\end{equation}
here $D_{-1}=D/0.1$, $\eta_{-1}=\eta/0.1$ and $\beta_{-1}=\beta/0.1$.
The remarkable property of disk dynamo supported BZ is undependable from mass of the progenitor.
It is mostly depend only from SMS angular momentum (\ref{DJ}), Higher angular momentum -- longer and brighter event (see fig.\ref{mdot} left panel).

The total power of such event is enormous ${E_{tot} \approx t_d\times  \dot{E}_{BZ} \sim 10^{56}}$ ergs.
This energy is enough to sweep away the protogalaxy with baryonic mass $\sim 10^8 \; M_{\odot}$ 

Taking in to account jet collimation, we can expect very bright sources, which could be detected
from very large cosmological distance.

\section{Discussion}
\label{Discussion}

SMS can be borne only in very early stages of star formation ($z \approx 10$), the observational time 
should be 
\begin{equation}
t_{do} = (1+z) t_d \approx 10 \mbox{ days} 
\label{tdo}
\end{equation}

Due to large z the spectra should be shifted to Hard X-Ray. If our analogy with short-hard GRB 
is right then we expect spectral peak in aria $60 keV$

We do not know for sure what is the central engine for short-hard gamma ray burst (GRB),
but it is most probable is merging of two compact stars, accretion disk formation and
collimation of the jet produced by strong wind from the disk. Here we may expect the similar behaviour.
The same collimation properties leads us to opening angle 0.1 rad. Put the factor dilution $A \sim 10^{-3}$

Let us estimate brightness of such burst. For ($z \approx 10$) and Cosmological parameters 
$\Omega_0=\Omega_{\Lambda}+\Omega_m=1$ and $\Omega_{\Lambda}=0.72$, $\Omega_{\Lambda}=0.28$ \citep{kom09}
luminosity distance will be $d_L\approx 3.2 \times 10^{29}$ cm \citep{muh05}, so flux on the earth should be
\begin{equation}
F = \frac{\dot{E}_{BZ}}{ 4\pi d_L^2 A} \approx 1.3\times 10^{-7} \eta_{-1} e_{c,-1} \beta_{-1} {D_{-1}}  \mbox{erg cm$^2$ s$^{-1}$}
\label{flux}
\end{equation}
here $e_{c,-1}=e_c/0.1$ is conversion coefficient of Blandford-Znaek flux to radiation. 
SMS were formed first in  gas clouds  with mass $\sim 10^8 M_{\odot}$. Follow the work \cite{BVR06} we can estimate the density of SMS at $z=10$ as ${n_{mh}\sim 0.01 Mpc^{-3}}$.

We can estimate the rate of such events in analogy to work \cite{KB09c}. 
Let us assume, for the sake of simplicity, that all supercollapsars go off 
simultaneously at cosmological time $t_e$ corresponding to $z=20$ (a moderate 
spread around this redshift will not significantly change the result). 
In flat Universe the observed time separation between events occurring 
simultaneously at $r_0$ and $r_0+dr_0$, where $r_0$ is the commoving radial 
coordinate, is $dt_o=cdr_0$. The corresponding physical volume within one steradian of 
the BAT's field of view, is 
$$
dV=a^3(t_e) r_0^2 dr_0,
$$ 
where $a(t_e)=(1+z)^{-1}$ is the scaling factor of the Universe at $t=t_e$  
(in the calculations we fix the scaling factor via the condition $a(t_o)=1$). 
$r_0$ and $t_e$ are related via $r_0=r\sub{L}(1+z)^{-1}$. Putting all this together 
we find the rate to be 
\begin{equation}
   f_c = {\cal A}\frac{ c n\sub{mh} r\sub{L}^2}{(1+z)^5} 
   \simeq 0.03 \, {\cal A}_{-3} \frac{n\sub{mh}}{10^{-2}} \mbox{yr}^{-1} .  
\label{fc}
\end{equation}     

This events should rear and the long term observations program could reveal them.

\section{Conclusions}
\label{conclusions}

In spite of the significant progress in the astrophysics of Gamma Ray Bursts, 
both observational and theoretical, it may still take quite a while before we 
fully understand both the physics of the bursts and the nature of their 
progenitors. At the moment there are several competing theories and too many 
unknowns. Similarly, we know very little about the star formation in the early 
Universe.  For this reason, the analysis presented above is rather speculative 
and the numbers it yields are not very reliable. Further efforts are required 
to develop a proper theory of supercollapsars and to make firm conclusions on 
their observational impact.
The collapse of SMS could be one of the most powerful event in the universe.
This event can destroy the seed cluster and form single SMBH. The expected 
very long duration of bursts and their relatively low brightness imply  that 
a dedicated search program using the image trigger will be required. Such 
search would be useful even in the case of non-detection as this would put 
important constraint on the models of early star formation, GRB progenitors, and SMBHs. 

\section*{Acknowledgments}
Author appreciated to Prof. S. Komissarov and F. Aharonyan for useful discussion.
This research was funded by PPARC under the rolling grant
``Theoretical Astrophysics in Leeds'' (MVB). 
We appreciate for partial support the NORDITA program 
on Physics of relativistic flows.

\end{document}